\documentstyle[12pt]{article}

\begin{document}

\input amssym.tex

\title{Discrete quantum modes of the Dirac field in $AdS_{d+1}$ 
backgrounds}

\author{Ion I. Cot\u aescu\\ {\it The West University of Timi\c soara,}\\{\it V. 
Parvan Ave. 4, RO-1900 Timi\c soara}}

\maketitle

\begin{abstract}

It is shown that the free Dirac equation in spherically symmetric static 
backgrounds of any dimensions can be put in a simple form using  
a special version of Cartesian gauge in Cartesian  coordinates. 
This is manifestly covariant under the transformations of the isometry group 
so that  the generalized spherical coordinates can be separated in terms of 
angular spinors like in the flat case, obtaining a pair of radial 
equations.  In this approach the equation of the free field Dirac in 
$AdS_{d+1}$ backgrounds is analytically solved obtaining the formula of the 
energy levels and the corresponding normalized eigenspinors. 

Pacs: 04.62.+v
\end{abstract}
\

\section{Introduction}


The interest in propagation of the quantum free fields on de Sitter or anti-de 
Sitter ($AdS$) spacetimes is due to the hope that these simple examples 
of matter fields minimally coupled with the gravitational field of the 
background could suggest new ways to further developments of the quantum field 
theory in curved spacetimes. Moreover, the discovery of the AdS/CFT 
correspondence \cite{M} brings in a central position the theory of the free 
fields on the $(d+1)$-dimensional $AdS$ spacetimes ($AdS_{d+1}$) since their 
quantum modes can be related to  the  conformal field theories in 
the Minkowski-like boundaries of the $AdS_{d+1}$ manifolds \cite{EW}.

The quantum modes of the scalar free fields in central charts of the 
$AdS_{d+1}$ backgrounds are well-studied either for $d=3$ \cite{AIS} or in  
the general case of any $d$ \cite{BL,C1}. However, for the Dirac
field we know only the solutions in the central charts of the $AdS_{3+1}$ 
spacetime we have obtained few years ago \cite{C2}. These were derived using a 
special Cartesian tetrad-gauge which points out the central symmetry as a 
manifest one \cite{ES} helping us to simply separate the spherical variables 
of the Dirac equation in terms of the angular spinors defined in special 
relativity \cite{TH}. Recently, Gu, Ma and  Dong have generalized the technique 
of separating spherical variables of the Dirac equation in Minkowski spacetime 
to flat (pseudo-Euclidean) spacetimes with generalized spherical coordinates 
of any  dimensions \cite{XYGu}. This offers us the opportunity to generalize 
our method to any $(d+1)$-dimensional curved spacetime with static 
charts having the $SO(d)$-symmetry, we call here (generalized) central charts. 
The idea is to write the Dirac equation in a suitable gauge allowing us to 
take over  the whole procedure of the separation of generalized spherical 
variables worked out for the flat manifolds with central potentials \cite{XYGu}. 
In this way we have to obtain simple radial equations that may be solved for 
particular cases as that of the $AdS_{d+1}$ spacetimes. The present article is 
devoted to this problem.   

The main point of our approach is the mentioned Cartesian gauge which 
can be generalized to central backgrounds of any dimensions, where it will 
play a similar role in showing off the manifest $SO(d)$ symmetry of the 
generalized central backgrounds. This leads to a simple covariant form of the 
Dirac equation in Cartesian coordinates and produces manifest covariant $SO(d)$ 
generators of the spinor representation whose orbital and spin parts commute 
to each other \cite{CML,ES}. In these conditions the separation of the 
generalized  spherical 
variables can be done in terms of the angular momentum eigenspinors defined 
in the $(d+1)$-dimensional flat backgrounds \cite{XYGu}. 
After the separation of the angular variables there remain only two 
radial wave functions for which we write down the radial equations in the 
general case  of any central chart with generalized spherical coordinates. 
These can be analytically solved for the $AdS_{d+1}$ backgrounds in the same 
manner as in the case of $d=3$ \cite{C2}, finding only discrete quantum modes.

We start in the second section outlining a convenient version of 
gauge-covariant Dirac theory in any dimensions. In Sec.3  we define the 
Cartesian gauge in Cartesian  coordinates and we bring the Dirac equation in a 
simpler form, called {\em reduced} equation, which has similar properties as 
that in flat spacetime, including the discrete symmetries. The next section is 
devoted to the separation of variables in generalized spherical coordinates 
which generates the radial equations. In Sec.5 we obtain the quantum modes of 
the Dirac field in the central charts with generalized spherical coordinates of 
the $AdS_{d+1}$ spacetime, deriving the normalized energy eigenspinors and 
the energy levels. The parity and charge-conjugation transformations we use are 
defined in Appendix.

\section{The gauge-covariant Dirac theory in $(d+1)$ dimensions}
\

The theory of the Dirac field in curved spacetimes of any dimensions involves 
three basic ingredients that can be chosen in different ways. These are the 
local chart (i.e., natural frame), the gauge fields defining the local frames 
and the Clifford algebra. In these conditions, the physical  meaning of the 
whole theory remains independent on the choice of these elements only if we 
assume that this is gauge-covariant. On the other hand, the generalization to 
a larger number of dimensions may be physically relevant in the sense  of 
the Kaluza-Klein theories only if the extra-dimensions are space-like. For this 
reason we consider here the gauge-covariant theory of the Dirac spinors in 
backgrounds with $d+1$ dimensions among them only one is time-like.   

We start with a such pseudo-Riemannian manifold, $M_{d+1}$, 
whose pseudo-Euclidean model has the flat metric 
$\eta$ of signature $(1,d)$, corresponding to the 
gauge group $G(\eta)=SO(1,d)$. Since the Dirac theory involves local 
orthogonal (non-holonomic) frames, we choose in $M_{d+1}$ a local chart  
with coordinates $x^{\mu}$, $\alpha,...,\mu,\nu,...=0,1,2,...,d$, 
and introduce local frames using the gauge fields $e_{\hat\alpha}(x)$ and 
$\hat e^{\hat\alpha}(x)$ labeled by local (hated) 
indices, $\hat\alpha,...,\hat\mu,\hat\nu,...=0,1,2,...,d$. 
Their components accomplish 
\begin{equation}  
e_{\hat\alpha}^{\mu}\hat e_{\nu}^{\hat\alpha}=\delta_{\mu}^{\nu}\,,\quad
e_{\hat\alpha}^{\mu}\hat e_{\mu}^{\hat\beta}=\delta_{\hat\alpha}^{\hat\beta}\,,
\quad 
g_{\mu\nu} e_{\hat\alpha}^{\mu} e^{\nu}_{\hat\beta}=
\eta_{\hat\alpha \hat\beta}\,, 
\end{equation}
and give the components of the metric 
tensor of $M_{d+1}$ in the natural frame,  
$g_{\mu\nu}(x)=\eta_{\hat\alpha \hat\beta}
\hat e ^{\hat\alpha}_{\mu}(x) \hat e^{\hat\beta}_{\nu}(x)$.    

Bearing in mind that the irreducible representations of the Clifford algebra 
have only an odd number of dimensions, we consider a $(2m+1)$-dimensional 
Clifford algebra with $2m\ge d$, acting on a spinor space with $2^m$ dimensions. 
From the basis of this algebra we choose a suitable set of $d+1$ 
$\gamma$-matrices and match the phase factors in order to have   
\begin{equation}\label{ACOM}
\{ \gamma^{\hat\alpha},\, \gamma^{\hat\beta} \} 
=2\eta^{\hat\alpha \hat\beta}\,. 
\end{equation}
Then  $(\gamma^0)^{\dagger}=\gamma^{0}$ and  $(\gamma^i)^{\dagger}=-\gamma^{i}$  
($i,j,k,...=1,2,...,d$) which means that  the Dirac adjoint of any matrix $A$ 
can be defined in usual manner as 
$\overline{A}=\gamma^{0}A^{\dagger}\gamma^{0}$ so 
that the $\gamma$-matrices be self-adjoint,  
$\overline{\gamma^{\hat\alpha}}=\gamma^{\hat\alpha}$.  
Notice that  the point-dependent matrices 
$\gamma^{\mu}(x)=e^{\mu}_{\hat\alpha}(x)\gamma^{\hat\alpha}$  are also 
self-adjoint with respect to the Dirac adjoint.

The group $G(\eta)$ admits an universal covering group $\tilde G(\eta)$ that is 
simply connected and has the same Lie algebra. The spin operators 
\begin{equation}\label{SAB} 
S^{\hat\alpha \hat\beta}=\frac{i}{4}\left[\gamma^{\hat\alpha},\,
\gamma^{\hat\beta}\right]
\end{equation}
represent the basis-generators of the {\em spinor} representation of 
$\tilde G (\eta)$ which, in general, can be reducible. 
The real valued parameters of ${\tilde G}(\eta)$ are the  components   
$\omega_{\hat\alpha\hat\beta}=-\omega_{\hat\alpha\hat\beta}$ of the 
skew-symmetric tensors, $\omega$ defining the operators 
\begin{equation}\label{TeS}
T(\omega)=e^{-iS(\omega)}\,,\quad S(\omega)=\frac{1}{2}
\omega_{\hat\alpha\hat\beta} S^{\hat\alpha\hat\beta}\,,
\end{equation}
which transform the gamma-matrices according to the rule
\begin{equation}\label{TgT}
[T(\omega)]^{-1}\gamma^{\hat\alpha}T(\omega)=\Lambda^{\hat\alpha\,\cdot}
_{\cdot\,\hat\beta}(\omega)\gamma^{\hat\beta} 
\end{equation}
where
\begin{equation}\label{Lam}
\Lambda^{\hat\alpha\,\cdot}_{\cdot\,\hat\beta}(\omega)=
\delta^{\hat\alpha}_{\hat\beta}
+\omega^{\hat\alpha\,\cdot}_{\cdot\,\hat\beta}
+\frac{1}{2}\,\omega^{\hat\alpha\,\cdot}_{\cdot\,\hat\lambda}
\omega^{\hat\lambda\,\cdot}_{\cdot\,\hat\beta}+...
+\frac{1}{n!}\,\underbrace{\omega^{\hat\alpha\,\cdot}_{\cdot\,\hat\lambda}
\omega^{\hat\lambda\,\cdot}_{\cdot\, \hat\sigma}
...\omega^{\hat\tau\,\cdot}_{\cdot\,\hat\beta}}_{n}+...\,,
\end{equation}
is a transformation of $G(\eta)$. We specify that the operators 
$T(\omega)\in {\tilde G}(\eta)$ are unitary with respect to the Dirac adjoint 
satisfying $\overline{T(\omega)}=[T(\omega)]^{-1}$.

Let us denote by $\psi$ the Dirac free field of mass $m$, and by 
$\overline{\psi}=\psi^{\dagger}\gamma^0$ its Dirac adjoint. In natural units 
(with $\hbar=c=1$) its gauge invariant action \cite{BD} is     
\begin{equation}\label{(action)}
{\cal S}[\psi,e]=\int\, d^{d}x\sqrt{g}\left\{
\frac{i}{2}[\overline{\psi}\gamma^{\mu}D_{\mu}\psi-
(\overline{D_{\mu}\psi})\gamma^{\mu}\psi] - 
m\bar{\psi}\psi\right\}
\end{equation}
where $g=|{\rm det}(g_{\mu\nu})|$ and 
$D_{\mu}=\nabla_{\mu}+\Gamma^{spin}_{\mu}$
are the covariant derivatives of the spinor field  
formed by the usual covariant derivatives $\nabla_{\mu}$ (acting in natural 
indices) and  the spin connection 
$\Gamma_{\mu}^{spin}=\frac{i}{2}
e^{\beta}_{\hat\nu}
(\hat e^{\hat\sigma}_{\alpha}\Gamma^{\alpha}_{\beta\mu}-
\hat e^{\hat\sigma}_{\beta,\mu} )
S^{\hat\nu\,\cdot}_{\cdot\,\hat\sigma}$.
The action of these covariant derivatives in the spinor sector is  
$D_{\mu}\psi=(\partial_{\mu}+\Gamma_{\mu}^{spin})\psi$. Since 
the spin connection satisfies 
 $\overline{\Gamma}_{\mu}^{spin}=-\Gamma_{\mu}^{spin}$,  
the quantity $\overline{\psi}\psi$ can be derived as 
a scalar, i.e. $\nabla_{\mu}(\overline{\psi}\psi)=
\overline{D_{\mu}\psi}\,\psi+\overline{\psi}\,D_{\mu}\psi=
\partial_{\mu}(\overline{\psi}\psi)$,
while the quantities $\overline{\psi}\gamma^{\alpha}\gamma^{\beta}...\psi$ 
behave as tensors of different ranks. Moreover, the use of covariant 
derivatives assures the covariance of the whole theory under the  gauge 
transformations 
\begin{eqnarray}
\hat e^{\hat\alpha}_{\mu}(x)&\to& \hat e'^{\hat\alpha}_{\mu}(x)=
\Lambda^{\hat\alpha\,\cdot}_{\cdot\,\hat\beta}[\omega(x)]
\,\hat e^{\hat\beta}_{\mu}(x)\,,\nonumber\\
e_{\hat\alpha}^{\mu}(x)&\to&  {e'}_{\hat\alpha}^{\mu}(x)=
\Lambda_{\hat\alpha\,\cdot}^{\cdot\,\hat\beta}[\omega(x)]
\,e_{\hat\beta}^{\mu}(x)\label{gauge}\,,\\
\psi(x)&\to&~\psi'(x)=T[\omega(x)]\,\psi(x)\,,\nonumber
\end{eqnarray}
due to the point-dependent parameters  $\omega_{\hat\alpha\hat\beta}(x)$, 
in the sense that ${\cal S}(\psi',e')={\cal S}(\psi, e)$. 
Thus we  reproduced  the main features of the familiar tetrad 
gauge-covariant theories with spin in  four dimensions from which we can take 
over now all the results arising from similar formulas. For example, in this 
way we find that 
\begin{eqnarray}
&&D_{\mu}(\gamma^{\nu}\psi)=\gamma^{\nu}D_{\mu}\psi\,, 
\label{Nabg}\\
&&[D_{\mu},\,D_{\nu}]\psi=
\textstyle{\frac{1}{4}}R_{\alpha\beta \mu\nu}
\gamma^{\alpha}\gamma^{\beta}\psi\,,
\end{eqnarray}
where $R$ is the Riemannian-Christoffel curvature tensor of $M_{d+1}$. 
The Dirac equation derived from the action (\ref{(action)}), 
${\cal E}_D\psi=m\psi$, involves the Dirac operator that can be 
written as  
\begin{equation}\label{(dd)}
{\cal E}_D= i\gamma^{\mu}D_{\mu}=
i\gamma^{\hat\alpha}e_{\hat\alpha}^{\mu}\partial_{\mu}
+ \frac{i}{2} \frac{1}{\sqrt{g}}\partial_{\mu}(\sqrt{g}e_{\hat\alpha}^{\mu})
\gamma^{\hat\alpha}
-\frac{1}{4}
e_{\hat\alpha}^{\mu}\{\gamma^{\hat\alpha},\,  \Gamma^{spin}_{\mu} \}\,.
\end{equation}
It is not difficult to verify that this is self-adjoint with respect to 
the Dirac adjoint.

A crucial problem is to find the generators of symmetries at the level of the 
relativistic quantum mechanics since these operators must commutes with the 
Dirac one. In the absence of other interactions  these are produced only by the 
 symmetries of the background. For explaining this mechanism  
it is worth reviewing some previous results we obtained in four 
dimensions and can be generalized to arbitrary dimensions. In \cite{ES} we 
defined the external symmetry group, $S(M)$, of a given manifold $M$ as the 
universal covering group of the isometry group, $I(M)$, pointing out that for 
each matter field there exists a specific representation of $S(M)$, 
{\em induced} by a linear representation of ${\tilde G}(\eta)$, that leaves 
the field equation invariant. Consequently, the generators of this 
representation are operators which commute with that of the field equation. 
They can be calculated starting with the Killing vectors corresponding to the 
isometries of $I(M)$ according to a rule  obtained by Carter and McLenaghan 
for the Dirac field \cite{CML} which states that for any Killing vector 
$k^{\mu}$ there exists an operator
\begin{equation}\label{Xk}
{X}_k=-ik^{\mu}D_{\mu}+\frac{1}{2}k_{\mu;\nu}e^{\mu}_{\hat\alpha}
e^{\nu}_{\hat\beta}S^{\hat\alpha\hat\beta}
\end{equation}
which commutes with ${\cal E}_D$. We have shown that these operators are just 
the generators of the representation of $S(M)$ in the space of the spinors 
$\psi$ that is induced by the spinor representation of $\tilde G(\eta)$. Each 
generator $(\ref{Xk})$ can be divided in an usual orbital part and a spin part, 
involving the spin matrices, whose forms depend on the choice of the gauge 
fields. When the spin parts commute with the orbital ones we say that the 
representation is manifest covariant. These results can be generalized for any 
$M_{d+1}$ manifold considering that the operators $S^{\hat\alpha\hat\beta}$ are  
those defined by Eq.(\ref{SAB}). 
 
In other respects, from the conservation of the electric charge, one can 
deduce that when $e^{0}_{i}=0$,  $i=1,2,...,d$, the relativistic scalar 
product of two spinors \cite{BD},
\begin{equation}\label{(sp)}
\left<\psi_{1},\psi_{2}\right>=\int_{D}d^{d}x\,\mu(x)\overline{\psi}_{1}(x)
\gamma^{0}\psi_{2}(x)\,, \quad 
\end{equation}
defined on the $d$-dimensional space domain $D$ of the chart we use, has 
the weight function
\begin{equation}\label{(weight)}
\mu(x)=\sqrt{g(x)}\,e_{0}^{0}(x)\,.
\end{equation}
The coherence of the whole theory is guaranteed by the fact that the 
conserved quantity associated to the Killing vector $k^{\mu}$, given 
by the Noether theorem, is the expectation value 
$\left<\psi, X_k \psi\right>$ of the operator (\ref{Xk}) calculated using 
the scalar product (\ref{(sp)}) \cite{C3}.     

\section{The Dirac equation in central backgrounds}

In what follows we shall focus only on the static central manifolds. These 
have static central charts with Cartesian coordinates, the time $x^0=t$ and 
space Cartesian coordinates $x^i$, $i=1,2,...,d$, where the metric is 
time-independent and spherically symmetric. The isometry groups of these 
manifolds, $I(M_{d+1})$, include as a subgroup the group of the static central 
symmetry, $I_c=T(1)\otimes SO(d)$, formed by time translations and 
$d$-dimensional orthogonal transformations of the Cartesian space coordinates 
seen as the components of the vector ${\bf x}\equiv(x^1,\,x^2,...,\,x^d)$.
The central symmetry requires the metric tensor, $g({\bf x})$, to transform 
manifestly covariant under the linear transformations of the space coordinates,
\begin{equation}\label{(rot)}  
x^i\to x'^{i}= R_{ij}x^{j}\,, \quad t'=t\,,
\end{equation}
produced by any $R\in SO(d)$. In these conditions the corresponding line 
element has the general form
\begin{equation}\label{(metr)} 
ds^{2}=g_{\mu\nu}({\bf x})dx^{\mu}dx^{\nu}=A(r)dt^{2}-[B(r)\delta_{ij}
+C(r)x^{i}x^{j}]dx^{i}dx^{j}
\end{equation} 
where  $A$, $B$ and $C$ are arbitrary functions of the Euclidean norm  
of ${\bf x}$, $r=|{\bf x}|=\sqrt{x^i x^i}$, which is invariant under $SO(d)$ 
transformations. In applications it is convenient to replace these functions 
by  new ones,  $u$,  $v$ and $w$, such that
\begin{equation}\label{(ABC)}   
A=w^{2}, \quad B=\frac{w^2}{v^2}, \quad 
C=\frac{w^2}{r^2}\left( \frac{1}{u^2}-\frac{1}{v^2}\right)\,.
\end{equation}
Then the metric appears as the conformal transformation of the simpler one 
with $w=1$.

In the case of the four-dimensional central backgrounds we proposed a  
Cartesian gauge in which the gauge fields are static and covariantly transform 
under  transformations similar to (\ref{(rot)}). We adopt the same type of gauge here, 
in the Cartesian charts of the central manifolds $M_{d+1}$, requiring the 
1-forms $d\hat x^{\hat\sigma}=\hat e_{\mu}^{\hat\sigma}dx^{\mu}$ to transform  
according to the rule
\begin{equation}\label{(tr)}
d\hat x ^{\hat\mu}\to d\hat x'^{\hat\mu}=\hat e^{\hat\mu}_{\alpha}(x')dx'^{\alpha}
=(Rd\hat x)^{\hat\mu}.
\end{equation}
Then the gauge fields must have components of the form 
\begin{eqnarray}
\hat e^{0}_{0}&=&\hat a(r), \quad \hat e^{0}_{i}=\hat e^{i}_{0}=0, \quad
\hat e^{i}_{j}=\hat b(r)\delta_{ij}+\hat c(r) x^{i}x^{j},\label{(eee)}\\
e^{0}_{0}&=& a(r), \quad  e^{0}_{i}= e^{i}_{0}=0, \quad
e^{i}_{j}= b(r)\delta_{ij}+ c(r) x^{i}x^{j},\label{(eee1)}
\end{eqnarray}
where, according to Eqs.(\ref{(metr)}) and (\ref{(ABC)}), we 
find that  
\begin{eqnarray}
\hat a&=&w, \quad \hat b=\frac{w}{v}, \quad \hat c=\frac{1}{r^2}
\left( \frac{w}{u}-\frac{w}{v}\right), \label{(abc)}\\
a&=& \frac{1}{w}, \quad  b=\frac{v}{w}, \quad  c=\frac{1}{r^2}
\left( \frac{u}{w}-\frac{v}{w}\right)\,.\label{(abc1)}
\end{eqnarray}                                                                                                                                                                                                                                                       
Since in this chart  
\begin{equation}
\sqrt{g}=B^{\frac{d-1}{2}}[A(B+r^{2}C)]^{\frac{1}{2}}=\frac{1}{ab^{d-1}(b+r^{2}c)}
=\frac{w^{d+1}}{uv^{d-1}}\,,
\end{equation}
we obtain the weight function (\ref{(weight)}) in our Cartesian gauge,  
\begin{equation}\label{(mu)}
\mu
=\frac{1}{b^{d-1}(b+r^{2}c)}
=\frac{w^d}{uv^{d-1}}\,.
\end{equation} 
From Eqs.(\ref{(abc)}) and (\ref{(abc1)}) we observe that $w$ must be 
positively defined in order to keep the same sense of the time axes of the 
natural and local frames. In addition, it is convenient to consider that the 
function $u$ is positively defined  and to try to define the radial 
coordinate $r$ so that $u=1$. However, the function $v$ can be of  
any sign. When $v>0$  then the space axes of the local frame at ${\bf x}=0$ 
are parallel with those of the natural frame, while for $v<0$ these are 
antiparallel.

Let us derive now the concrete form of the operators of the Dirac theory 
in this gauge. First we observe that if we replace the tetrad components in 
Eq.(\ref{(dd)}) then the last term of its left-hand side does not contribute. 
The remaining equation can be put in a simpler form introducing the 
{\it reduced} Dirac field, $\tilde\psi$, defined as
\begin{equation}\label{(cfu)}
\psi(x)=\chi(r)\tilde\psi(x)\,,
\end{equation}
where 
\begin{equation}\label{(chi)}
\chi=[\sqrt{g}(b+r^{2}c)]^{-1/2}=v^{\frac{d-1}{2}}w^{-\frac{d}{2}}\,.     
\end{equation}
After this substitution  we obtain the reduced Dirac equation, 
$\tilde {\cal E}_D\tilde \psi=m\tilde\psi$, in which the reduced Dirac 
operator,
\begin{equation}\label{(red)}
\tilde{\cal E}_D=i a(r)\gamma^{0}\partial_{t} + i b(r)\gamma^i 
\partial_i+
i c(r)(\gamma^i x^i)\left(\textstyle\frac{d-1}{2} +
x^i\partial_i\right)\,,
\end{equation}
is independent on the derivatives of  $a,\,b$ and $c$. 

We observe that the reduced Dirac equation has a manifest symmetry similar to 
that of the Dirac 
equation in flat backgrounds with central potentials. This is the consequence 
of our version of Cartesian gauge which produces usual linear representation   
of the external symmetry group $S_c(M_{d+1})$, associated with $I_c(M_{d+1})$, 
showing off the central symmetry in a manifest covariant form. Indeed, taking 
into account that $S_c(M_{d+1})=T(1)\otimes \tilde{SO}(d)$ involves the group  
$\tilde{SO}(d)\subset \tilde G(\eta)$ which is the universal covering group of 
$SO(d)$, we conclude that the spin operators $S^{ij}$ defined by Eq.(\ref{SAB}) 
are the basis-generators of the spinor representation of $\tilde{SO}(d)$. 
Furthermore, following the method of Ref.\cite{ES}, we find that 
the basis-generators of the induced representation of $S_c(M_{d+1})$ 
given by Eq.(\ref{Xk})  are the $T(1)$ generator, $i\partial_t$, and the 
$\tilde{SO}(d)$ ones, 
\begin{equation}\label{j}
{\cal J}_{ij}=L_{ij}+S_{ij}=-i(x^i\partial_j-x^j\partial_i)+S_{ij}\,,  
\end{equation}
which play the role of total angular momentum.
They commute with the operator (\ref{(red)}) and, obviously, have 
a manifest covariant form. 

On the other hand, the reduced Dirac equation can 
be put in Hamiltonian form $i\partial_t\tilde\psi={\cal H}\tilde\psi$ as in 
the case of the flat manifolds using the operators
\begin{eqnarray}
{\cal H}&=&-i\frac{u(r)}{r^2}(\gamma^0\gamma^i x^i)\left(\textstyle\frac{d-1}{2}+
x^i\partial_i\right)-i\frac{v(r)}{r^2}(\gamma^i x^i){\cal K}
+w(r)\gamma^0 m\,,\\
{\cal K}&=&\gamma^0 \left(S^{ij}L_{ij}+\textstyle\frac{d-1}{2}\right)\,.
\end{eqnarray}
Since $\gamma^0$ commutes with ${\cal J}_{ij}$ and ${\cal K}$ we have
\begin{equation}
\left[{\cal H},\, {\cal J}_{ij}\right]=0\,, \quad
\left[{\cal H},\, {\cal K}\right]=0\,.
\end{equation}
Consequently, all the properties related to the conservation of 
the angular momentum, including the separation of  variables in spherical 
coordinates, will be similar to those of the usual  Dirac theory in the 
flat spacetimes with $d+1$ dimensions \cite{XYGu}. Moreover, we note that the 
discrete transformations of parity and charge conjugation can be also defined. 
These leave Eq.(\ref{(red)}) invariant and have the same significance as in 
special relativity \cite{SW1,TH}. Thus, for example, the charge conjugation 
transforms each particular solution of positive frequency into the 
corresponding one of negative frequency (see the Appendix).    

\section{Separation of spherical coordinates}

The next step is to introduce the generalized spherical coordinates, 
$r$, $\theta_1, \theta_2,..., \theta_{d-1} $, \cite{T} associated with  
the space  coordinates of our natural Cartesian frame, 
\begin{eqnarray}
x^1&=&r\cos\theta_1\sin\theta_2...\sin\theta_{d-1}\,,\nonumber\\
x^2&=&r\sin\theta_1\sin\theta_2...\sin\theta_{d-1}\,,\label{sferic}\\
&\vdots&\nonumber\\
x^d&=&r\cos\theta_{d-1}\,.\nonumber
\end{eqnarray}
Then from Eqs.(\ref{(metr)}) and (\ref{(ABC)}) one obtains the line element   
\begin{equation}\label{(muvw)}
ds^{2}=w^{2}dt^{2}-\frac{w^2}{u^2}dr^2-
\frac{w^2}{v^2}r^2 d\theta^{2}\,,
\end{equation}
where $d\theta^{2}$ is the usual line element on the sphere $S^{d-1}$
\cite{T}, 
\begin{equation}
d\theta^2={d\theta_1}^2+\sin^2\theta_1 {d\theta_2}^2+
\sin^2\theta_1 \sin^2\theta_2{d\theta_3}^2 +\cdots\,.
\end{equation}
Consequently, in this chart we have 
\begin{equation}
\sqrt{g_s} =\sqrt{g}\,r^{d-1}\prod_{a=1}^{d-1}(\sin\,\theta_a)^{a-1}\,,
\end{equation}
and similarly for the weight function (\ref{(weight)}).
The special form of the reduced Dirac equation allows one to separate the 
spherical variables as in the case of the central motion in flat spacetimes, 
using the angular spinors $\phi_{\kappa,(j)}(\hat{\bf x})$ defined in 
Ref.\cite{XYGu}. These depend only on the spherical variables represented by 
the unit vector $\hat{\bf x}={\bf x}/r$, being determined by a set of weights 
$(j)$ of an irreducible representation of $\tilde{SO}(d)$ and the eigenvalue 
$\kappa=\pm |\kappa|$ of the operator ${\cal K}$ which concentrates all the 
angular operators of ${\cal H}$. 

The separation procedure is complicated 
being different for even or odd values of $d$ \cite{XYGu}. For odd $d$ the  
particular solution of given energy, $E$, and positive frequency corresponding 
to the irreducible representation $(j)$ may have the form
\begin{equation}\label{(psol)}
\tilde\psi_{E,\kappa,(j)}(t,{\bf x})=r^{-\frac{d-1}{2}}e^{-iEt}\left(
\begin{array}{c}
f^{+}_{E,\kappa}(r)\phi_{\kappa,(j)}(\hat{\bf x})\\
i f^{-}_{E,\kappa}(r)\phi_{-\kappa,(j)}(\hat{\bf x})
\end{array}\right)\,,
\end{equation}
where $\kappa$ can take  positive or negative  values. However, if $d$ is 
even then 
there are pairs of associated irreducible representations, $(j_1)$ and $(j_2)$, 
giving the same eigenvalue of the first Casimir operator, among them one takes 
the first one for positive values of $\kappa$ and  the second one for the 
negative values, $\kappa=-|\kappa|$. Consequently, the particular solutions 
read
\begin{eqnarray}
\tilde\psi_{E,|\kappa|,(j_1)}(t,{\bf x})&=&r^{-\frac{d-1}{2}}e^{-iEt}\nonumber\\
&&\times \left[f^{+}_{E,|\kappa|}(r)\phi_{|\kappa|,(j_1)}(\hat{\bf x})
+if^{-}_{E,-|\kappa|}(r)\phi_{-|\kappa|,(j_1)}(\hat{\bf x})\right]\,,
\label{sol1}\\
\tilde\psi_{E,-|\kappa|,(j_2)}(t,{\bf x})&=&r^{-\frac{d-1}{2}}e^{-iEt}\nonumber\\
&&\times \left[f^{+}_{E,-|\kappa|}(r)\phi_{-|\kappa|,(j_2)}(\hat{\bf x})
+if^{-}_{E,|\kappa|}(r)\phi_{-|\kappa|,(j_2)}(\hat{\bf x})\right]\,.
\label{sol2}
\end{eqnarray}
It is remarkable that both these types of solutions involve  the same 
set of radial wave functions  that depend on $E$ and   
\begin{equation}\label{kappa}
\kappa=\pm(\textstyle\frac{d-1}{2}+l)\,,\quad l=0,1,2,... \,,  
\end{equation}
where $l$ is an auxiliary orbital quantum number defining the 
representations $(j)$, $(j_1)$ and $(j_2)$ \cite{XYGu}. The radial 
wave functions obey a pair of radial equations similar to those found in the 
case of $d=3$ \cite{C2}. For a given  $\kappa$, we embed these radial equations 
in the eigenvalue problem   
\begin{equation}\label{(hfef)}
H_{\kappa}\,f_{E,\kappa}=E\,f_{E,\kappa}
\end{equation}   
of the {\em radial} Hamiltonian operator 
\begin{equation}
H_{\kappa}=\left(\begin{array}{cc}
    mw& u\frac{\textstyle d}{\textstyle dr}+\kappa\frac{\textstyle v}
{\textstyle r}\\
       -u\frac{\textstyle d}{\textstyle dr}+\kappa\frac{\textstyle v}
{\textstyle r}& -mw
\end{array}\right)
\end{equation}
acting in the two-dimensional space with elements  $f=(f^{+}, f^{-})^{T}$. 
After the separation of variables the scalar product (\ref{(sp)}) splits in 
angular and radial terms. Supposing that the angular spinors are normalized 
with respect to an angular scalar product, we remain with the  radial scalar 
product   
\begin{equation}\label{(spp)}
\left<\psi_{1},\psi_{2}\right>=
\left<f_{1}, f_{2}\right>=\int_{D_{r}}\frac{dr}{u}
{f}_{1}^{\dagger}{f}_{2}\,.
\end{equation}
The radial weight function $\mu \chi^{2}=1/u$, resulted from Eqs.(\ref{(mu)}) 
and (\ref{(chi)}), is just that we need in order to have 
$(u\partial_{r})^{\dagger}=-u\partial_{r}$ such that $H_{\kappa}$ be 
Hermitian  with respect to the scalar product (\ref{(spp)}). 
Thus we obtain an independent radial problem  which has to be solved in  
each particular case separately using appropriate methods.

\section{Quantum modes in $AdS_{d+1}$ spacetimes}

The $AdS_{d+1}$ spacetime is the  hyperboloid $\eta_{AB}Z^{A}Z^{B}=R^{2}$ 
of radius $R=1/\omega$ in the $(d+2)$-dimensional flat spacetime of 
coordinates $Z^{-1},\, Z^{0},\, Z^{1},...,\, Z^{d}$ and metric
$\eta_{AB}={\rm diag}(1,1,-1,...,-1)$, $A,\,B=-1,0,1,...d$.  Here 
we consider the static  chart of Cartesian coordinates  $(t, {\bf x})$ 
defined as 
\begin{eqnarray}
Z^{-1}&=&R\sec \omega r \cos \omega t\,,\nonumber \\
Z^{0}&=&R\sec \omega r \sin \omega t \,, \\ 
{\bf Z}&=&R\frac{{\bf x}}{r}\tan \omega r \,, \nonumber
\end{eqnarray}  
and the corresponding static chart with generalized spherical coordinates 
(\ref{sferic}) where the line element reads \cite{AIS,BL}
\begin{equation}\label{(leads)}
ds^{2}=\eta_{AB}dZ^{A}dZ^{B}=\sec^{2}\omega r\left(dt^{2}-dr^{2} -
\frac{1}{\omega^2}\sin^{2}\omega r\, d\theta^{2}\right)\,.
\end{equation}
In this chart  $r\in D_{r}=[0,\, \pi/2\omega)$ and, therefore, the whole space 
domain is $D=D_{r}\times S^{d-1}$. In addition, we specify that the time of 
$AdS_{d+1}$ must satisfy   
$t\in [-\pi/\omega,\pi/\omega)$ while  $t\in (-\infty,\infty)$ 
defines the universal covering spacetime of $AdS_{d+1}$ ($CAdS_{d+1}$) 
\cite{AIS}.

Furthermore, from Eq.(\ref{(leads)}) we identify
\begin{equation}\label{uvwADS}
u(r)=1\,,\quad w(r)=\sec \omega r\,, \quad v(r)=\omega r \csc \omega r\,,
\end{equation}
and, introducing the notation $k=m/\omega$ (i.e. 
$mc^{2}/\hbar\omega$ in usual units), we find the Hamiltonian
of the radial problem
\begin{equation}
H_{\kappa}=\left(\begin{array}{cc}
    \omega k\sec \omega r& \frac{\textstyle d}{\textstyle dr}+
\omega\kappa\csc\omega r\\
      - \frac{\textstyle d}{\textstyle dr}+\omega\kappa\csc\omega r
& -\omega k\sec \omega r
\end{array}\right)\,.
\end{equation}
Its form suggests us to perform the local rotation  $f\to \hat{f}=U(r)\,f=
(\hat f^{+},\,\hat f^{-})^T$ where   
\begin{equation}\label{(uder)} 
U(r)=\left(\begin{array}{cc}
    \cos \frac{\omega r}{2}&\sin 
\frac{\omega r}{2}\\
-\sin \frac{\omega r}{2}&\cos 
\frac{\omega r}{2}
\end{array}\right)\,.
\end{equation}
The transformed  Hamiltonian,   
\begin{equation}\label{(newh)}
\hat H_{\kappa} =U(r)H_{\kappa}U^{\dagger}(r)-\frac{\omega}{2} 1_{2\times 2}
=\left(\begin{array}{cc}
    \nu& \frac{\textstyle d}{\textstyle dr}+W_{\kappa}\\
      - \frac{\textstyle d}{\textstyle dr}+W_{\kappa}& -\nu
\end{array}\right)
\end{equation}
giving the new eigenvalue problem
\begin{equation}\label{(trrp)}
\hat H_{\kappa} \hat{f}=\left(E-\frac{\omega}{2}\right)\hat{f}\,,
\end{equation} 
has supersymmetry since it has the requested specific form  
with  $\nu=\omega(k+\kappa)$  and the superpotential
\begin{equation}\label{(super)}
W_{\kappa}(r)=\omega(\kappa\cot\omega r - k\tan \omega r)\,. 
\end{equation}
Consequently, the transformed radial wave functions, $\hat f^{\pm}$, 
satisfy the second order equations 
\begin{equation}
\left(-\frac{1}{\omega^2}\frac{d^2}{dr^2}+\frac{k(k\mp 1)}{\cos^{2}\omega r}+
\frac{\kappa(\kappa\mp 1)}{\sin^{2}\omega r}\right)
\hat f^{(\pm)}(r)=\epsilon^{2}\hat f^{(\pm)}(r)\label{(od1)}\,,\\
\end{equation}
where  $\epsilon=E/\omega-1/2$. The general solutions can be written in terms 
of Gauss hypergeometric functions \cite{AS},   
\begin{eqnarray}
\hat f^{\pm}(r)&=&N_{\pm}
\sin^{2s_{\pm}}\omega r\cos^{2p_{\pm}}\omega r \label{(gsol)}\\
&&\times F\left(s_{\pm}+p_{\pm}-\frac{\epsilon}{2},
s_{\pm}+p_{\pm}+\frac{\epsilon}{2}, 2s_{\pm}+\frac{1}{2}, \sin^{2}\omega r
\right)\,,\nonumber
\end{eqnarray}
depending on  parameters that must satisfy 
\begin{eqnarray}
2s_{\pm}(2s_{\pm}-1)&=&\kappa(\kappa\mp 1)\,,\label{(2s)}\\ 
2p_{\pm}(2p_{\pm}-1)&=&k(k\mp 1)\,,\label{(2p)} 
\end{eqnarray}
and on the normalization factors $N_{\pm}$.  
The next step is to select the suitable values 
of these parameters  and to calculate $N_{+}/N_{-}$ such that the functions 
$\hat f^{\pm}$ should be solutions of the transformed radial problem 
(\ref{(trrp)}) with  a good physical meaning. This can be achieved only 
when $F$ is a polynomial selected by a suitable quantization condition 
since otherwise $F$ is strongly divergent for $\sin^{2}\omega r\to 1$. 
Then the  functions $\hat f^{\pm}$ will be square integrable 
with  normalization factors calculated according to the condition
\begin{equation}\label{(norm)}
\left<{f},{f}\right>= 
\left<\hat{f},\hat{f}\right>
=\int_{D_{r}}dr\,\left( |\hat f^{+}(r)|^{2}+
|\hat f^{-}(r)|^{2}\right)=1\,, 
\end{equation}  
resulted from the fact that the matrix (\ref{(uder)}) is orthogonal.

The discrete energy spectrum is given by the  particle-like $CAdS$ quantization 
conditions
\begin{equation}\label{(quant)}
\epsilon=2 (n_{\pm}+s_{\pm}+p_{\pm})\,, \quad \epsilon>0\,,
\end{equation}
that must be compatible with each other, i.e.
\begin{equation}\label{(comp)}
n_{+}+s_{+}+p_{+}=n_{-}+s_{-}+p_{-}\,.
\end{equation}
Hereby we see  that there is only one independent {\em radial} 
quantum number, $n_{r}=0,1,2,...$. In other respects, if we express  
the solutions (\ref{(gsol)}) in terms of Jacobi 
polynomials, we observe that these functions remain square integrable for 
$2s_{\pm}>-1/2$ and $2p_{\pm}>-1/2$. Since from Eq.(\ref{kappa}) we have 
$|\kappa|\ge \frac{d-1}{2}$, we are forced to consider only  the positive 
solutions of Eqs.(\ref{(2s)}). 
However, the Eqs.(\ref{(2p)}) admit either  
the solutions $2p_{+}=k$ and $2p_{-}=k+1$ defining the boundary 
conditions of {\em regular} modes or other possible values,     
$2p_{+}=-k+1$ and $2p_{-}=-k$, giving the {\em irregular} modes for which 
$k<1/2$ \cite{C2}. Here we restrict ourselves  
to write down only the energy eigenspinors of the regular modes on $CAdS$.   

Let us consider first $\kappa=|\kappa|$ and choose $2s_{+}=\kappa$ and 
$2s_{-}=\kappa+1$ for which  Eq.(\ref{(comp)}) is accomplished provided 
$n_{+}=n_{r}$ and $n_{-}=n_{r}-1$. Then the functions $\hat f^{\pm}$ given by 
Eqs.(\ref{(gsol)}) represent a solution of the transformed radial problem 
(\ref{(trrp)}) only if
\begin{equation}\label{(npnm)}
\frac{N_{-}}{N_{+}}=\frac{2n_{r}}{2|\kappa|+1}\,.
\end{equation} 
These can be expressed  in terms of Jacobi polynomials as 
\begin{eqnarray}
\hat f^{+}_{n_{r},|\kappa|}(r)&=& 
N \sin^{|\kappa|}\omega r \cos^{k}\omega r 
P_{n_{r}}^{(|\kappa|-\frac{1}{2},k-\frac{1}{2})}(\cos 2\omega r)\,,
\label{(1)}\\
\hat f^{-}_{n_{r},|\kappa|}(r)&=&N 
\sin^{|\kappa|+1}\omega r \cos^{k+1}\omega r 
P_{n_{r}-1}^{(|\kappa|+\frac{1}{2},k+\frac{1}{2})}(\cos 2\omega r)\,,
\nonumber
\end{eqnarray}
We specify  that from Eq.(\ref{(npnm)}) we understand that the second equation 
of (\ref{(1)}) gives $\hat f^{-}=0$ when $n_{r}=0$. 
For $\kappa=-|\kappa|$ we use the same procedure finding that 
$2s_{+}=|\kappa|+1$, $2s_{-}=|\kappa|$, $n_{+}=n_{-}=n_{r}$ and
\begin{equation}
\frac{N_{-}}{N_{+}}=-\frac{2|\kappa|+1}{2n_{r}+2k+1}\,.
\end{equation} 
The corresponding normalized radial wave functions are 
\begin{eqnarray}
\hat f^{+}_{n_{r},-|\kappa|}(r)&=&N\left[\frac{n_{r}+k+\frac{1}{2}}{n_{r}+
|\kappa|+\frac{1}{2}}\right]^{\frac{1}{2}}
\nonumber\\
&&\times \sin^{|\kappa|+1}\omega r \cos^{k}\omega r 
P_{n_{r}}^{(|\kappa|+\frac{1}{2},k-\frac{1}{2})}(\cos 2\omega r)\,,
\label{(2)}\\
\hat f^{-}_{n_{r},-|\kappa|}(r)&=&
-N \left[\frac{n_{r}+|\kappa|+\frac{1}{2}}{n_{r}+k+\frac{1}{2}}
\right]^{\frac{1}{2}}
\nonumber\\
&&\times\sin^{|\kappa|}\omega r \cos^{k+1}\omega r 
P_{n_{r}}^{(|\kappa|-\frac{1}{2},k+\frac{1}{2})}(\cos 2\omega r)\,.
\nonumber
\end{eqnarray}
In both cases, the general normalization factor calculated from  
Eq.(\ref{(norm)}) is 
\begin{equation}
N=\sqrt{2\omega}\left[\frac{n_{r}!\,\Gamma(n_{r}+k+|\kappa|+1)}
{\Gamma(n_{r}+|\kappa|+\frac{1}{2})\Gamma(n_{r}+k+\frac{1}{2})}\right]^{\frac{1}{2}}
\,.
\end{equation} 

Finally, using the inverse transformation of (\ref{(uder)}) we  obtain 
the radial wave functions 
\begin{equation}
f^{\pm}_{n_r,\kappa}(r)=\hat f^{\pm}_{n_r,\kappa}(r)\cos\frac{\omega r}{2} 
\mp \hat f^{\mp}_{n_r,\kappa}(r)\sin\frac{\omega r}{2}
\end{equation}
of the particular solutions of positive frequency (\ref{(psol)}), (\ref{sol1}) 
and (\ref{sol2}) of the reduced Dirac equation. 
The last step is to restore the form of the field $\psi$, according to 
Eqs.(\ref{(cfu)}), (\ref{(chi)}) and (\ref{uvwADS}),  
writing down the particular solutions of positive frequency of the original 
Dirac equation,
\begin{equation}
\psi_{n_r,\kappa,(j)}^{pa}(t,{\bf x}) = \left( \frac{\omega r}{\sin\omega r}
\right)^{\frac{d-1}{2}}\cos^{\frac{d}{2}}\omega r\,\tilde\psi_{n_r,\kappa,(j)}
(t,{\bf x})\,.
\end{equation}
If these are interpreted as particle-like solutions then 
the antiparticle-like ones can be derived directly through the 
charge conjugation as 
\begin{equation}
\psi_{n_r,\kappa,(j)}^{ap} 
=\left(\psi_{n_r,\kappa,(j)}^{pa}\right)^c 
=C\left(\overline{\psi}_{n_r,\kappa,(j)}^{\,pa}\right)^T \,, 
\end{equation}
where the matrix $C$ is defined by Eq.(\ref{CCo}). 

The energy levels result from Eq.(\ref{(quant)}) where we must take 
into account that $\omega k=m$ and $\omega\epsilon=E-\omega/2$. Thus we 
obtain 
\begin{equation}\label{Enrk}
E_{n_r,\kappa}=\left\{
\begin{array}{lcl}
m+\omega(2n_{r}+\kappa+\frac{1}{2})&{\rm for}&\kappa=|\kappa|\\
m+\omega(2n_{r}+|\kappa|+\frac{3}{2})&{\rm for}&\kappa=-|\kappa|
\end{array} 
\right.
\end{equation}
which suggests us to introduce the principal quantum number 
\begin{equation}\label{princ}
n=\left\{
\begin{array}{lcl}
2n_{r}+|\kappa|-\frac{d-1}{2}=2n_r +l&{\rm for}&\kappa=|\kappa|\\
2n_{r}+|\kappa|-\frac{d-1}{2}+1=2n_r+l+1&{\rm for}&\kappa=-|\kappa|
\end{array} 
\right.
\end{equation}
taking the values $0,1,2,....$ since $l$ ranges as in 
Eq.(\ref{kappa}) and $n_r=0,1,2,...$. With its help we can write   
the compact formula of the energy levels
\begin{equation}\label{(enlev)}
E_{n}=m+\omega\left(n+\frac{d}{2}\right)\,,\quad n=0,1,2,....
\end{equation} 
which is similar to that of the $d$-dimensional homogeneous harmonic oscillator 
but having a relativistic rest energy. These energy levels are deeply 
degenerated  and the problem of determining the degree of degeneracy  
seems to be complicated requiring a special study.  

\section{Conclusions}

We demonstrated that our Cartesian gauge in central charts leads to simple 
reduced Dirac equations in Cartesian coordinates and similar radial problems 
in spherical coordinates for any central background $M_{d+1}$. 
These reduce to a pair of radial equations that depend on $d$ only 
through the quantum number $\kappa$. This property may allow one to 
solve the radial problem  for large sets of spacetimes of the same type 
and to study how depend the quantum modes on $d$.

In the case of the  $CAdS_{d+1}$ spacetimes we found that the energy spectrum 
of the Dirac field minimally coupled with gravity is discrete and equidistant 
like that of the scalar field but the ground state energy is quite different. 
We remind the reader that the energy levels of the scalar free field are 
\cite{BL}
\begin{equation}
E_n^{sc} =\sqrt{m^2+\frac{d^2}{4}\omega^2}+\omega\left(n+\frac{d}{2}\right)
\,,\quad n=0,1,2,... ~. 
\end{equation}
The difference is due to the fact that in curved spacetimes the free Dirac 
equation is no more the square root of the Klein-Gordon equation with the 
same mass. This gives rise to new problems concerning the physical 
interpretation of the mass terms in theories where the matter fields are 
minimally coupled with gravity. In other respects, our result has consequences 
even in holography \cite{EW} since this indicates that the conformal dimensions 
of the boundary Klein-Gordon and Dirac (or Majorana) fields of the ADS/CFT 
conjecture may be different.

\subsection*{Acknowledgments}

I would like to thank E. Papp for useful discussions and drawing me attention on 
Ref.\cite{XYGu}.

\appendix

\section{Parity and charge conjugation}

In the theory of the Dirac field in curved spacetimes $M_{d+1}$ the discrete 
transformations can be defined in a similar way as in special relativity. 
Thus the parity changes ${\bf x}\to -{\bf x}$ and $\psi({\bf x},t)\to \gamma^0
\psi(-{\bf x},t)$ leaving the reduced Dirac equation invariant.  
The form of the charge conjugation depends on the representation of the 
Dirac matrices. These satisfy Eqs.(\ref{ACOM}) and, consequently, they must be 
either symmetric or skew-symmetric.  Let us assume that there are $s+1$ 
symmetric $\gamma$-matrices, namely $\gamma^0$ and $s$ matrices with 
space-like indices,
$\gamma^{\hat\alpha_1},\,\gamma^{\hat\alpha_2},...,\gamma^{\hat\alpha_s}$.
Then the matrix
\begin{equation}\label{CCo}
C=(-1)^{\frac{s}{2}}\gamma^0\gamma^{\hat\alpha_1}\gamma^{\hat\alpha_2},...,
\gamma^{\hat\alpha_s}
\end{equation}
has the properties $C^{-1}=C^{T}=(-1)^{s}\overline{C}$ and 
$C\gamma^{\hat\mu}C^{-1}=(-1)^s (\gamma^{\hat\mu})^T$. With its help one can 
define the charge conjugated spinor 
$\psi^{c}=C{\overline{\psi}}^{\,T}$ of the spinor $\psi$ and verify that the 
reduced Dirac equation given by the operator (\ref{(red)}) remains invariant 
under the charge conjugation  $\psi\to \psi^c$. Note that this transformation 
is point-independent which suggests that the vacuum state could be stable (or 
invariant \cite{BD}) in quantum field theories based on field equations 
invariant under this type of charge-conjugation. Particularly, it is known 
that the Euclidean vacuum of the $AdS$ spacetime is invariant \cite{AIS,BD}.

\end{document}